\documentclass[%
reprint,
onecolumn,
amsmath,amssymb,
aps,
prl,
superscriptaddress,
]{revtex4-2}

\usepackage{graphicx}% Include figure files
\usepackage{dcolumn}% Align table columns on decimal point
\usepackage{bm}% bold math
\usepackage{color}
\usepackage{braket}

\begin{document}
	
%\preprint{APS/123-QED}

\title{Near-Degenerate Quadrature-Squeezed Vacuum Generation on a Silicon-Nitride Chip}

\author{Yun Zhao}
\email{yz3019@columbia.edu}
\affiliation{Department of Electrical Engineering, Columbia University, New York, NY 10027, USA}
\author{Yoshitomo Okawachi}
\affiliation{Department of Applied Physics and Applied Mathematics, Columbia University, New York, NY 10027, USA}
\author{Jae K. Jang}
\affiliation{Department of Applied Physics and Applied Mathematics, Columbia University, New York, NY 10027, USA}
\author{Xingchen Ji}
\affiliation{Department of Electrical Engineering, Columbia University, New York, NY 10027, USA}
\author{Michal Lipson}
\affiliation{Department of Electrical Engineering, Columbia University, New York, NY 10027, USA}
\affiliation{Department of Applied Physics and Applied Mathematics, Columbia University, New York, NY 10027, USA}
\author{Alexander L. Gaeta}
\affiliation{Department of Electrical Engineering, Columbia University, New York, NY 10027, USA}
\affiliation{Department of Applied Physics and Applied Mathematics, Columbia University, New York, NY 10027, USA}

\date{\today}% It is always \today, today,

\begin{abstract}
	Squeezed states are a primary resource for continuous-variable (CV) quantum information processing. To implement CV protocols in a scalable and robust way, it is desirable to generate and manipulate squeezed states using an integrated photonics platform. In this Letter, we demonstrate the generation of quadrature-phase squeezed states in the radio-frequency carrier sideband using a small-footprint silicon-nitride microresonator with a dual-pumped four-wave-mixing process. We record a squeezed noise level of 1.34 dB ($\pm$0.16 dB) below the photocurrent shot noise, which corresponds to 3.09 dB ($\pm$0.49 dB) of quadrature squeezing on chip. We also show that it is critical to account for  the nonlinear behavior of the pump fields to properly predict the squeezing that can be generated in this system. This technology represents a significant step toward creating and manipulating large-scale CV cluster states that can be used for quantum information applications including universal quantum computing.
\end{abstract}

\maketitle

Quantum optics can revolutionize information acquisition and processing, including sensing \cite{Caves_PRD_1981,Giovannetti_NatPhot_2011}, communication \cite{Scarani_RMP_2009}, and computation \cite{Kok_RMP_2007}, by offering new paradigms to achieve performance that is superior to classical approaches. The implementation of such schemes requires the use of nonclassical states of light, where information can be carried by both discrete (DV) and continuous (CV) quantum variables. The quantum sources required to implement most DV and CV protocols are single photons and squeezed light, respectively. Compared to the DV information processing, CV schemes provide powerful alternatives with unique features, such as high efficiency state characterization and unconditional state manipulation \cite{Braunstein_RMP_2005}. Recently, the demand of scalability in quantum information processing has stimulated efforts to produce quantum light sources on a photonic chip. While fully integrated quasideterministic single photons have been used in proof-of-principle quantum information protocols \cite{Qiang_NatPhot_2018}, the on-chip generation of squeezed light that is suitable for CV protocols remains challenging and has shown less scalability due to the large feature sizes and long fabrication times in reported experiments \cite{Anderson_OL_1995, Serkland_OL_1995, Kanter_OE_2002, Mondain_PhotRes_2019, Lenzini_SciAdv_2018,Otterpohl_Optica_2019}.

Squeezed states represent a key resource for CV quantum information processing, with applications in universal quantum computing \cite{Lloyd_PRL_1999,Menicucci_PRL_2006}, quantum error correction \cite{Braunstein_PRL_1998(2),Lloyd_PRL_1998}, quantum teleportation \cite{Braunstein_PRL_1998}, quantum secret sharing \cite{Lau_PRA_2013} and quantum key distribution \cite{Gottesman_PRA_2001}. There are many ways to generate a quadrature-squeezed state optically, most of which are based on the three pioneering experiments reported during 1985 and 1986, namely, noncollinear degenerate four-wave mixing in atomic ensembles \cite{Slusher_PRL_1985}, self-phase modulation (SPM) in a $\chi^{(3)}$ medium \cite{Shelby_PRL_1986}, and degenerate parametric down-conversion (DPDC) in a $\chi^{(2)}$ medium \cite{Wu_PRL_1986}. It is worth noting that the nonlinear processes are degenerate for the carrier frequency, while noise squeezing arises from photon correlations at nearby frequency sidebands, which are nondegenerate. To date, most of the chip-based squeezed state generations rely on the DPDC process \cite{Anderson_OL_1995, Serkland_OL_1995, Kanter_OE_2002, Lenzini_SciAdv_2018, Mondain_PhotRes_2019}, however, the large-scale integration of and fabrication on $\chi^{(2)}$ materials remains challenging.

As an alternative, researchers have investigated the generation of squeezed states on a platform compatible with the standard complementary metal oxide semiconductor (CMOS) process on a silicon chip, which provides a scalable solution for CV quantum sources. This calls for a squeezing process with $\chi^{(3)}$ nonlinearity in a small-footprint structure. The first direct observation of photocurrent noise squeezing on a silicon chip was shown by Dutt \textit{et al.} in a silicon-nitride (SiN) microresonator \cite{Dutt_PRAppl_2015}, in which a spontaneous four-wave-mixing (SFWM) process was employed and a reduction was observed in the relative intensity fluctuations between bright signal and idler beams below the shot-noise level. Recently, Vaidya \textit{et al.} \cite{Vaidya_Arxiv_2019} successfully observed quadrature-phase squeezing with the SFWM scheme operating below oscillation threshold. However, the squeezed sidebands are generated from two largely separated (190 GHz) cavity modes and can only be measured via a bichromatic heterodyne measurement \cite{Marino_JOSAB_2007}. Many current CV protocols require single-spatiotemporal-mode squeezed states \cite{Vernon_PRAppl_2019}, which is incompatible with this two-mode squeezing scheme. An alternative on-chip squeezing scheme is demonstrated by Cernansky and Politi which uses SPM \cite{Cernansky_arXiv_2019}. Quadrature squeezing was observed for frequencies above 500 MHz. However, a large amount of excess noise was observed for lower frequencies due to their close proximity to the strong pump. 

\begin{figure*}
	\centering
	\includegraphics{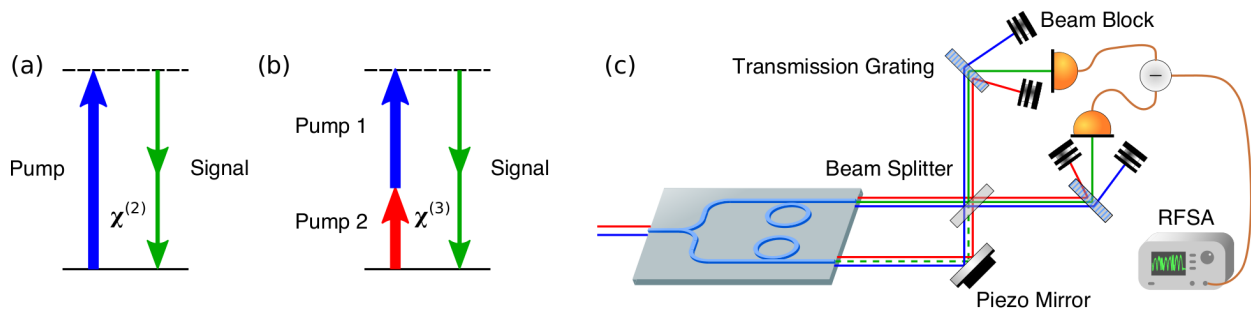}
	\caption{(a) The DPDC process used in $\chi^{(2)}$-based squeezing, and (b) the DPFWM process used in this experiment for on-silicon-chip squeezing. (c) Experimental schematic (not to scale). Two pump lasers at 1543 nm and 1559 nm are coupled onto the chip which simultaneously generates a squeezed state and a LO. The two output fields are overlapped on a 50/50 beam splitter with the arm length difference controlled by a piezo steering mirror. After the beam splitter, the pump fields are separated from the signal fields with two transmission gratings. The signal fields are then detected by a balanced photodetector, and the rf signal is recorded by a rf spectrum analyzer (RFSA). The full experiment schematic including pump preparation and stabilization is presented in the Supplementary Material, which includes Ref. \cite{Carmon_OE_2004}.}
	\label{figSetup}
\end{figure*}

In this Letter, we generate quadrature squeezing near the degenerate carrier frequency by employing a dual-pumped degenerate four-wave-mixing (DPFWM) scheme \cite{Okawachi_OL_2015}, in which two pump fields provide parametric gain for the degenerate signal field at a frequency centered between those of the pump fields [Fig. \ref{figSetup}(b)]. This scheme can be viewed as the $\chi^{(3)}$ counterpart to the DPDC process [Fig. \ref{figSetup}(a)], where the upper virtual transition level is driven by two pump fields rather than by one. Previously, this DPFWM scheme has been used for on-chip random number generation \cite{Okawachi_OL_2016}, degenerate photon-pair generation \cite{He_Optica_2015}, and boson sampling \cite{Paesani_NatPhys_2019}. The possibility of using it to generate squeezed states was recently studied theoretically by Vernon \textit{et al.} \cite{Vernon_PRAppl_2019}, but no experimental demonstration has been shown prior to this Letter.

Our on-chip squeezing device is based on the SiN-on-insulator platform, which has the advantage of extremely low propagation loss and large-scale monolithically integrability. It has been widely used in photonics research with both classical and quantum applications \cite{Moss_NatPhot_2013, Gaeta_NatPhot_2019,Kues_NatPhot_2019}. Particularly, its low propagation loss ($\approx$0.28 dB/cm for the current device and 0.08 dB/cm has been shown \cite{Ji_Optica_2017}) and ease of fabrication allow for implementation of cavity-enhanced squeezing schemes on chip, which greatly boosts the squeezing level and shrinks the device footprint. Quadrature-squeezed states are best characterized by homodyne measurements that are also essential to CV quantum information protocols. To perform a homodyne measurement, a local oscillator (LO) that is phase coherent with the squeezed state is required. In this Letter, we generate the squeezed state and a coherent LO on the same chip simultaneously. We fabricate two identical ring resonators that are evanescently coupled to their respective bus waveguides with the coupling gaps chosen such that one ring is critically coupled (i.e., round-trip coupling ratio = round-trip propagation loss ratio) and the other is overcoupled (i.e., round-trip coupling ratio $>$ round-trip propagation loss ratio). Consequently, the overcoupled ring has a higher oscillation threshold. For the current experiment, the overcoupled ring has an escape efficiency of 2/3 (i.e., loaded $Q$ is 3$\times$ that of the intrinsic $Q$) and the intrinsic $Q$ of the microresonators is measured to be 1.3 $\times 10^6$. Each SiN ring has a platinum microheater above the silicon dioxide cladding to adjust for any mismatch in resonance frequencies that arises from accumulated phase shifts in the coupling region and nonuniform material strains across the chip. Platinum is chosen for its CMOS-process compatibility, inert chemical properties, and high heat tolerance. The two bus waveguides split the pumps evenly through a multimode interferometric beam splitter (MMI) at the input. When the pump frequencies are tuned to the resonant modes of both microresonators, the critically coupled ring oscillates, producing a bright local oscillator, and the overcoupled ring remains below threshold, generating a squeezed state. The matching of the transverse-mode profiles between the LO and the squeezed state are naturally achieved by this scheme since they are generated in the same waveguide mode. Henceforth, we will refer to the critically coupled ring as the oscillator and the overcoupled ring as the squeezer. 

\begin{figure}
	\centering
	\includegraphics{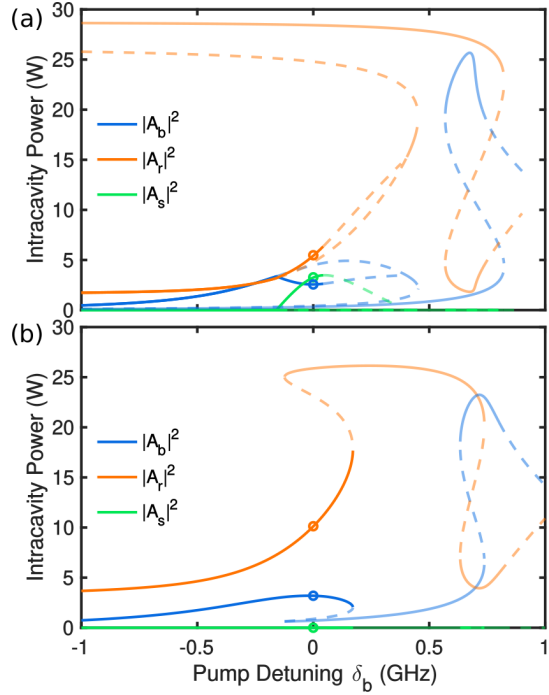}
	\caption{Numerical simulation of equilibrium states in a dual-pumped $\chi^{(3)}$ cavity. For clarity, some equilibrium states are not shown. Stable states are plotted in solid lines and unstable states in dashed lines. Dark lines indicate the branch accessed in the experiment and light lines indicate the branches to avoid. The states corresponding to the final experimental condition are shown in circles. The simulation parameters are chosen as $\delta_r=2\pi\times 650$ MHz, $\gamma = 1$ W$^{-1}$m$^{-1}$, L = 230 $\mu$m, $f_\delta = 2\pi\times 500$ GHz, $P_b$ = 48 mW, $P_r$ = 54 mW, $\alpha = 2\pi\times 150$ MHz, (a) $\theta = 2\pi\times 150$ MHz, and (b) $\theta = 2\pi\times 300$ MHz.}
	\label{figBifurcation}
\end{figure}

Two stable states with a $\pi$ phase difference are supported by the oscillator \cite{Graham_QS,Drummond_PRA_1989}. Each time the pump fields are turned on, the LO settles into one of two stable phases. However for the purpose of this experiment, the two states yield the same results since a $\pi$ phase shift of the LO results in a measurement on the same quadrature component. To ensure that degenerate oscillation is the dominant nonlinear process, we dispersion engineer the microresonators to have small normal group-velocity dispersion (GVD) at the degenerate frequency such that the parametric gain peaks at this frequency. More information concerning the design and dispersion engineering of this degenerate oscillator can be found in \cite{Okawachi_OL_2015}. The experimental characterization of the LO is shown in the Supplementary Material, which includes Refs. \cite{Okoshi_ElectronLett_1980,LeThomas_Optica_2018,Huang_PRA_2019,Ji_Optica_2017}.

A critical aspect for this dual-pumped system is that it can display rich nonlinear dynamics due to the existence of SPM and cross-phase modulation (XPM) \cite{DelBino_SciRep_2017}. To more fully understand the dynamics of each resonator, we model the pump and signal evolution with the same set of three-coupled-mode equations \cite{Chembo_PRA_2010},
\begin{align}
\frac{dA_b}{dt} &= i\gamma L f_\delta(|A_b|^2+2|A_r|^2+2|A_s|^2)A_b + i\gamma L f_\delta A_r^\ast A_s^2 \notag\\
&\quad\quad -(\frac{\alpha+\theta}{2} +i\delta_b)A_b + \sqrt{\theta f_\delta P_b},\\
\frac{dA_r}{dt} &= i\gamma L f_\delta(2|A_b|^2+|A_r|^2+2|A_s|^2)A_r + i\gamma L f_\delta A_b^\ast A_s^2 \notag\\ 
&\quad\quad -(\frac{\alpha+\theta}{2} + i\delta_r)A_r + \sqrt{\theta f_\delta P_r},\\
\frac{dA_s}{dt} &= i\gamma Lf_\delta(2|A_r|^2+2|A_b|^2+|A_s|^2)A_s + i2\gamma Lf_\delta A_s^\ast A_bA_r \notag\\
&\quad\quad  -(\frac{\alpha+\theta}{2} +i\delta_\beta)A_s,
\end{align}
where $A_i$ ($i$ = $b,r,s$) are the amplitudes of the short-wavelength (blue) pump, long-wavelength (red) pump, and the signal fields, respectively, $\alpha$ is the scattering loss rate, $\theta$ is the bus-ring coupling rate, $f_\delta$ is the free spectral range (FSR), $\gamma$ is the nonlinear coefficient, $L$ is the cavity length, $P_i$ ($i$ = $b,r$) are the input pump powers, and $\delta_i$ ($i$ = $b,r$) are the pump detunings with respect to the linear cavity resonances in angular frequencies. We choose the convention such that $\delta_i > $ 0 indicates red detuning. $\delta_\beta = \delta_b/2+\delta_r/2+2\pi^2\beta_2 Lm^2f_\delta^3$ is the signal detuning due to second-order dispersion, where $\beta_2$ is the GVD coefficient and $m$ is the mode index difference between the pump and the signal. We normalize the field amplitudes such that $|A_i|^2$ represents the average power in the cavity. The $\alpha$ and $\theta$ parameters can be directly related to the intrinsic and loaded $Q$ factors $Q_i$ and $Q_l$ as $\alpha = \omega/Q_i$ and $\alpha + \theta = \omega/Q_l$, where $\omega$ is the angular frequency of the signal field. The complex nonlinear dynamics can be illustrated by plotting the bifurcation behaviors of the system. We fix the detuning of the red pump and sweep the detuning of the blue pump while searching for all equilibrium states (Fig. \ref{figBifurcation}). The stable states are shown in solid lines and the unstable states are shown in dashed lines. In the oscillator ring, a signal field is generated with a small-detuned blue pump if the red pump is initially on the lower branch of the equilibrium states (darker lines). We show examples of other equilibrium states that are supported within the same detuning parameter space with lighter lines, which need to be avoided as they are unsuitable for either LO or squeezed state generation. The squeezer ring is identical to the oscillator ring except for a higher bus-ring coupling rate ($\theta$). Importantly, as their pumps are derived from the same laser sources, both the oscillator and the squeezer operate under the same detuning conditions. To generate a squeezed state, it is necessary to operate below oscillation threshold while keeping a high intracavity power of both pumps. By examining Fig. \ref{figBifurcation}(b), we find that it can be achieved by the same branch of equilibrium states as the oscillation branch in Fig. \ref{figBifurcation}(a). Experimentally, this branch of states can be accessed by tuning the red pump from a long wavelength to the desired detuning value (650 MHz red detuned) followed by tuning the blue pump from a short wavelength into the resonance. The experiment is performed at the detunings that have a LO power slightly lower than the maximum power (Supplementary Material). The corresponding equilibrium states are indicated by circles in Fig. \ref{figBifurcation}. Under nearly all conditions, the intracavity powers of the two pump fields are not equal even for equal input powers due to their tendency to repel each other at high powers \cite{DelBino_SciRep_2017}. This can make both the optical parametric oscillation and the squeezing processes less energy efficient but poses no fundamental limit to the attainable squeezing level.

\begin{figure}
	\centering
	\includegraphics{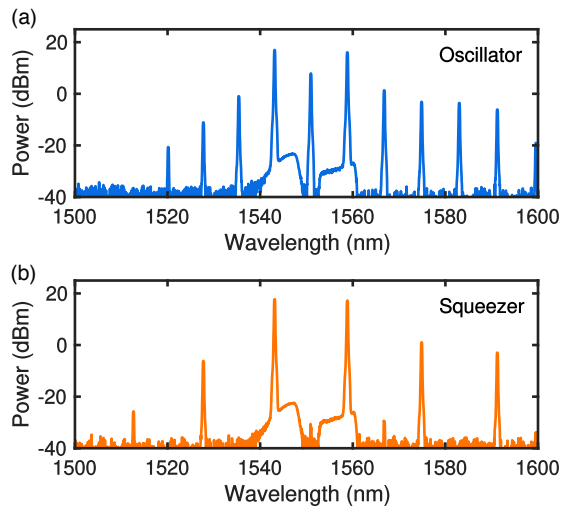}
	\caption{Output spectra from the oscillator (a) and the squeezer (b), respectively. The FSR of the devices is 4 nm (500 GHz).}
	\label{figOPO}
\end{figure}

The output spectrum of the LO ring is shown in Fig. \ref{figOPO}(a). The two pump fields are at 1543 nm and 1559 nm, respectively and the LO is generated at 1551 nm, two FSRs away from the pump resonances. We also observe frequency components generated by other four-wave-mixing FWM processes, but they are sufficiently weak and far detuned from the LO that they can be neglected. The LO light is easily separated from the other frequency components with a grating [Fig. \ref{figSetup}(c)] owing to the large FSR of microcavities. The spectrum corresponding to squeezed state generation is shown in Fig. \ref{figOPO}(b). The small peak at the signal wavelength (1551 nm) is due to the backreflection from the facets of waveguides ($\approx$ 4\% per facet) and the backscattering of the LO microresonator. This spurious signal is measured to be more than 40 dB lower than that of the LO. The effect of a weak seed laser on squeezing has been studied in the context of coherent control of squeezed states and has been found to be tolerable for low power seeds or high frequency detection sidebands \cite{McKenzie_PRL_2004,Goda_PRA_2005}. It is also possible to further eliminate this spurious signal by using angled waveguide facets or a filter ring before the squeezer. The other bright frequency components in Fig. \ref{figOPO}(b) are a result of stimulated FWM processes, whose effects are negligible for squeezed state generation near the degenerate point. 

\begin{figure}
	\centering
	\includegraphics{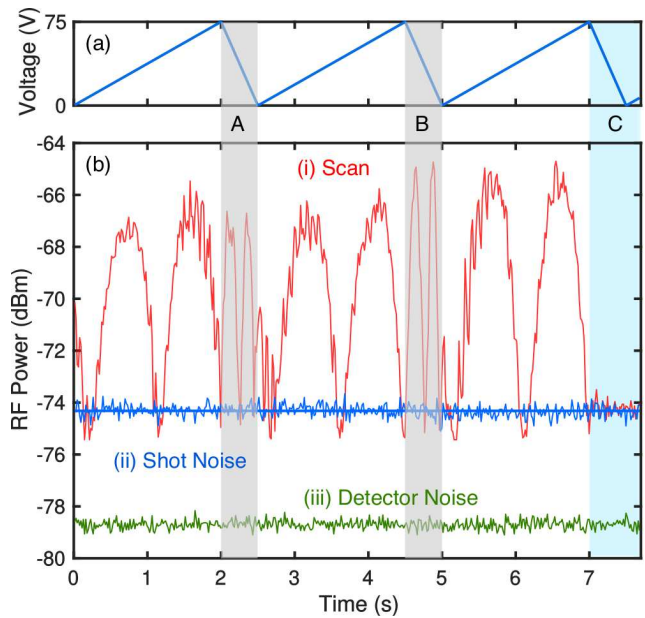}
	\caption{(a) Piezoscan voltage and (b) detected rf noise power while the delay is scanned, (i) with the squeezed state, (ii) without the squeezed state, and (iii) without light. Shaded regions $A$ and $B$ correspond to the return part of the scan. After the third scan, we block the squeezed state to record an extra copy of the shot noise, which is shown in shaded region $C$.}
	\label{figSqueezing}
\end{figure}

To characterize the squeezed state with a homodyne measurement [Fig. \ref{figSetup}(c)], we interfere the two output beams (including all frequency components) from the chip on a beam splitter (Thorlabs CCM5-BS018, $T$ = 42\%, $R$ = 47\%). We use two transmission gratings (LightSmyth T-966C-1610-90) to spatially separate the different frequency modes of the output, for which 97\% of the light goes into the first diffraction order. In principle, the beam splitter can be replaced by an on-chip evanescent coupler, and the gratings can be replaced by on-chip add-drop ring filters, both of which have negligible losses. The two LO beams after the gratings are collected into high numerical aperture fibers (Thorlabs FP200ERT) and detected by fiber-coupled balanced detectors (Thorlabs PDB150C). We use a polarizer (Thorlabs LPNIR050) to balance the DC photocurrents of the two photodiodes (PDs), where the initial imbalance arises from the imperfect splitting ratio of the beam splitter and the unequal responsivities of the PDs. We use a piezosteering mirror on the squeezer path to control the relative delay between the squeezed state and the LO. A maximum scanning voltage of 75 V corresponds to a phase change of $\approx$ 2.6$\pi$. Figure \ref{figSqueezing} shows the photocurrent noise for three scanning periods at a sideband frequency of 40 MHz, with a resolution bandwidth of 100 kHz and a video bandwidth of 100 Hz. At the end of the third scan, we block the squeezed state and record the shot-noise level. For a clear comparison, we record a second copy of the shot noise for a same time span as the scan with the squeezed state blocked, which shows excellent agreement with the first shot-noise measurement, which indicates that the generated LO is stable. In Fig. \ref{figSqueezing}(b), the scanned signal dips below the shot-noise level, which indicates squeezing. The high fluctuation level of the antisqueezed quadrature indicates that the squeezer is operated close to oscillation threshold. From Fig. \ref{figSqueezing}(b), we obtain a directly observed squeezing level of 0.81 dB $\pm$ 0.09 dB, where the uncertainty is taken as the standard error of the measurement (Supplementary Material). The low detector dark-noise clearance also artificially reduces the measured squeezing level. This effect can be removed by a simple numerical subtraction, which yields a detector noise corrected squeezing level of 1.34 dB $\pm$ 0.16 dB. The measured squeezing level is also reduced by the losses in the measurement system. We estimate the total loss to be 48\%, including Fresnel reflection losses, mirror losses, beam splitter loss, grating losses and the detector quantum efficiency. After correcting for these losses, we infer an on-chip squeezing level of 3.09 dB $\pm$ 0.49 dB.

Similar to squeezing in a $\chi^{(2)}$ cavity, the level of squeezing from our $\chi^{(3)}$ microresonator is ultimately limited by the cavity scattering losses and the output coupling ratio. However, due to the proximity between the pump fields and the signal in our scheme, other parasitic $\chi^{(3)}$ processes can occur. The nonlinear processes of SPM and XPM modify the cavity detuning of the squeezed signal but do not change the maximum squeezing level that is achieved near oscillation threshold. The more deleterious processes are SFWM and FWM Bragg scattering processes \cite{Vernon_PRAppl_2019} that couple the squeezed mode to other cavity modes, which degrades the tight field correlation between the upper and lower frequency sidebands of the squeezed mode. Indeed, a theoretical model including all parasitic processes (see Supplementary Material, which includes Refs. \cite{Vernon_PRAppl_2019,Collett_PRA_1984,Vernon_PRA_2015,Zhao_Optica_2020,Little_JLT_1997}) shows that no more than 0.8 dB of squeezing can be generated if these processes are unmitigated. The operation regime (circles in Fig. \ref{figBifurcation}) in this experiment is chosen to suppress the parasitic processes based on their different phase-matching conditions compared to the DPFWM process. We present a detailed analysis in the Supplementary Material, where we show that the pump state in Fig. \ref{figBifurcation}(b) leads to a theoretical squeezing level of 3.5 dB, agreeing with the experimental observation.

In conclusion, we experimentally demonstrate the first silicon-chip-based generation of quadrature-phase squeezed states in the rf carrier sideband. This also represents the first time photocurrent noise squeezing is observed from the DPFWM process. We also show that the pump waves can exhibit rich nonlinear dynamical behavior that determines the squeezing levels. The SiN platform we used for squeezing is a mature technology for large-scale fabrication of linear optical components which, combined with squeezed states, can be used to form entangled CV states \cite{Masada_NatPhot_2015} and CV cluster states. More details about combining the on-chip squeezed states with currently available integrated beamsplitters and detectors for quantum information processing can be found in the Supplementary Material, which includes Refs. \cite{Huh_NatPhot_2015,Menicucci_PRL_2014,Fukui_PRX_2018,Carolan_Science_2015,Calkins_OE_2013,Raffaelli_QST_2018,Porto_JOSAB_2018,Yu_OE_2018,Piprek_SPIE_2003,Lin_PhotRes_2017,Vahlbruch_PRL_2016,Oser_arXiv_2020,Baragiola_PRL_2019}. We believe this technology represents a significant step towards the fully on-chip implementation of CV quantum protocols and possibly photonic-based universal quantum computers.

We thank Professor Imad Agha, Dr. Chaitanya Joshi, Dr. Avik Dutt, Chaitali Joshi, and Bok Young Kim for helpful discussions. This work was supported by Army Research Office (ARO) (Grant No. W911NF-17-1-0016), National Science Foundation (NSF) (Grant No. CCF-1640108, EFMA-1641094), Semiconductor Research Corporation (SRC) (Grant No. SRS 2016-EP-2693-A), and Air Force Office of Scientific Research (AFOSR) (Grant No. FA9550-15-1-0303). This work was performed in part at the Cornell Nano-Scale Facility, which is a member of the National Nanotechnology Infrastructure Network, supported by the NSF, and at the CUNY Advanced Science Research Center NanoFabrication Facility.

During the preparation of this manuscript, another study on squeezed state generation based on the SiN platform was reported \cite{Zhang_arXiv_2020}.


%%%%%%%%%% Merge with supplemental materials %%%%%%%%%%
\pagebreak
\widetext
\begin{center}
	\textbf{\large Near-degenerate quadrature-squeezed vacuum generation on a silicon-nitride chip: supplementary material}
\end{center}
%%%%%%%%%% Merge with supplemental materials %%%%%%%%%%
%%%%%%%%%% Prefix a "S" to all equations, figures, tables and reset the counter %%%%%%%%%%
\setcounter{equation}{0}
\setcounter{figure}{0}
\setcounter{table}{0}
\setcounter{page}{1}
\setcounter{section}{0}
\makeatletter
\renewcommand{\theequation}{S\arabic{equation}}
\renewcommand{\thefigure}{S\arabic{figure}}
\renewcommand{\bibnumfmt}[1]{[S#1]}
\renewcommand{\citenumfont}[1]{S#1}
%%%%%%%%%% Prefix a "S" to all equations, figures, tables and reset the counter %%%%%%%%%%

\section{Experiment setup}
\begin{figure*}[b]
	\centering
	\includegraphics{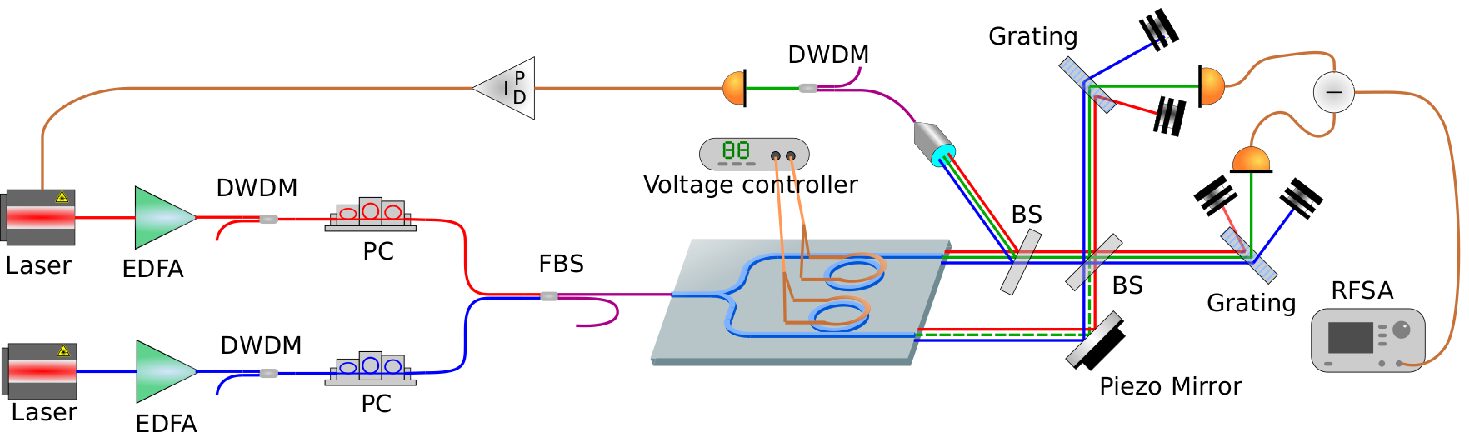}
	\caption{Experimental setup. EDFA, erbium-doped fiber amplifier; DWDM, dense wavelength-division multiplexer; PC, polarization controller; FBS, fiber beam splitter; BS, beam splitter; RFSA, radio frequency spectrum analyzer.}
	\label{figSetupDetail}
\end{figure*}
The pump source consists of two external-cavity diode lasers (ECDL's, New Focus Velocity$^{\text{TM}}$ TLB-6728) at 1543 and 1559 nm, respectively. Both lasers are amplified by erbium-doped fiber amplifiers (EDFA's) and sent to dense wavelength division multiplexers (DWDM's) to filter the amplified spontaneous emission (ASE) noise. We insert polarization controllers (PC's) after the DWDM's and combine the two lasers with a 50/50 beam splitter. The combined pumps are coupled into the chip (TE$_{00}$ mode) using a lensed fiber. We connect a two-channel voltage source to the two integrated microheaters through a multi-contact wedge. The heaters are used to match the resonances of the microrings through thermal tuning.

The photonic chip consists of a multimode-interferometric beam splitter (MMI) and two microring resonators. The splitting ratio of the MMI is 1.04 :1 for all relevant wavelengths (1543 nm to 1559 nm). Both microrings have cross sections of 730 nm $\times$ 1050 nm and radii of 46 $\mu$m. The coupling gaps are 575 nm (critically coupled) and 525 nm (overcoupled), respectively. The critically coupled ring serves as the oscillator and the overcoupled ring serve as the squeezer. The pump powers for the microrings can be measured at the outputs of each bus waveguide, which yields 48 mW for the 1543 nm pump and 54 mW for the 1559 nm pump.

Due to the existence of self- (SPM) and cross-phase modulation (XPM), the dual pumped optical parametric oscillators (OPO's) exhibit extremely rich nonlinear dynamics. Our theoretical model shows that many branches of steady states can be found. In this experiment we only explore one branch of the states that we can most reliably access. Moreover, this operation regime provides significant suppression of parasitic nonlinear processes (see section 3). First, we set the 1559 nm pump on the red side of its corresponding resonance with a detuning of $\approx$ 2 GHz. Then we tune the 1543-nm pump from the blue side into resonance. As the 1543-nm pump is tuned, the resonance around 1559 nm redshifts due to both XPM and thermal refractive effect, effectively reducing the detuning of the 1559-nm pump. With proper detuning of the 1543 nm pump, we find a regime where the oscillator generates a bright LO while the squeezer is below threshold. With this scheme, the LO power is sensitive to the frequency separation of the two pumps, where a pump frequency drift of $\approx$ 100 MHz can noticeably change the power of the LO. To improve the stability of the generation scheme, we tap a small portion of the LO power and use it as a reference to stabilize the frequency of the 1543 nm laser through a proportional-integral-derivative (PID) controller. The 1559 nm pump is left free running as its slight drift would be compensated by the thermal refractive effect, known as soft thermal locking \cite{Carmon_OE_2004}.

At the output of the photonic chip, we use an aspheric lens (not shown in the figure) to collimate the outputs of the two bus wavegudes. We then overlap the two output beams on a 50/50 beam splitter. We place a piezo steering mirror on the squeezer arm to actively control the phase difference between the two arms. After the beam splitter we use two transmission gratings to separate the signal from the pumps. Finally we couple the two signal beams into two uncoated large-numerical-aperture fibers which are connected to a balanced photodetector. The electrical signal is sent to a radio frequency spectrum analyzer (RFSA) for noise characterization.

\section{Local oscillator noise characterization}
\begin{figure*}
	\centering
	\includegraphics{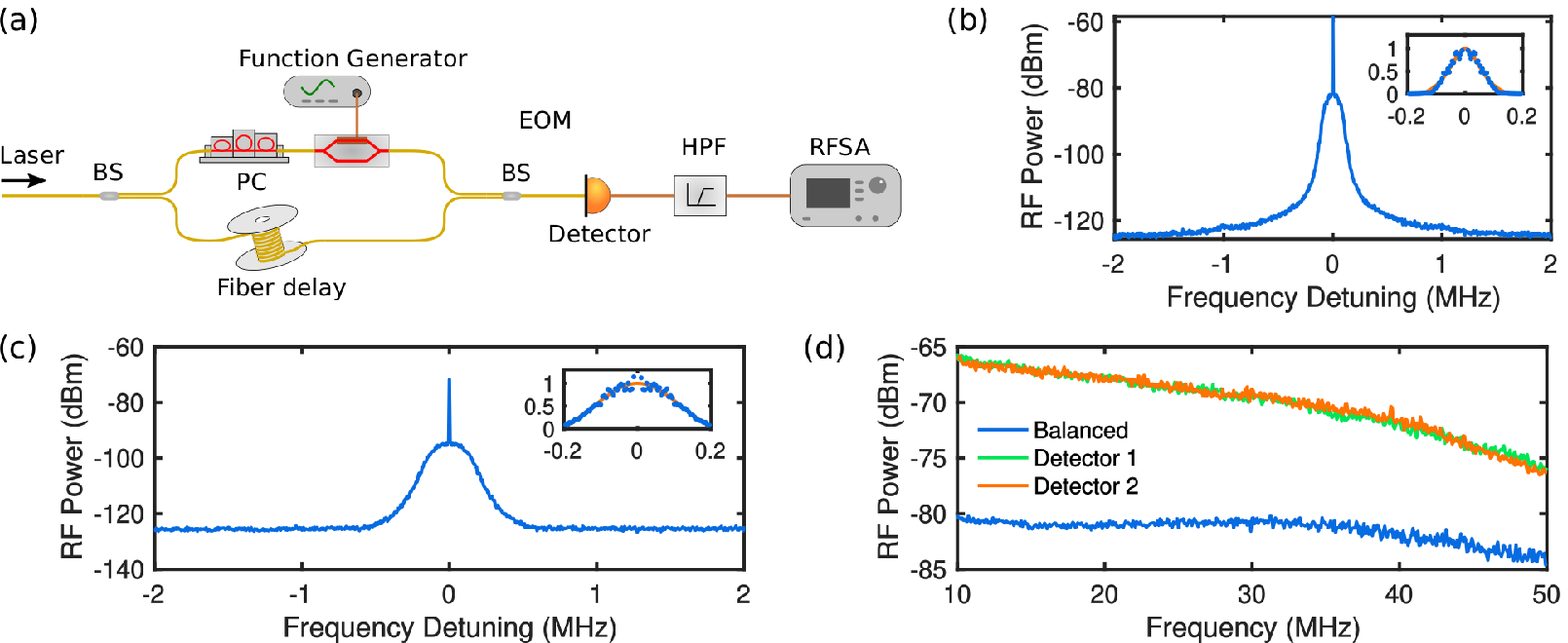}
	\caption{(a) Experimental setup for delayed self-heterodyne measurement. BS, fiber based beam splitter; PC, polarization controller; EOM, electro-optical modulator; HPF, electrical high pass filter; RFSA, radio-frequency spectrum analyzer. (b, c) Frequency noise measurement of the pump laser (b) and the LO beam (c). Insets, Lorentzian fit (red) of the noise envelope. (d) LO beam intensity noise measurement.}
	\label{figSelfheterodyne}
\end{figure*}
We characterize the phase noise of the LO using a delayed self-heterodyne technique \cite{Okoshi_ElectronLett_1980}. We couple the grating-separated signal beam into a fiber (not shown in Fig \ref{figSelfheterodyne}) and use a 50/50 beam splitter to split the signal into two parts. The top arm is intensity modulated at 1 GHz to create a frequency sideband. The bottom arm is delayed by a pathlength significantly longer than the expected coherence length (15 km of single-mode fiber). The two arms are subsequently recombined and the beat signal is detected by a photodiode. We use a high pass filter (HPF) to reject the beatings near the zero frequency and retain only the signal near the modulation sideband. The beatnote is recorded by a RFSA. 

We show the measured sideband signal in Fig. \ref{figSelfheterodyne}(b) and (c), where the sharp peak corresponds to the modulation RF tone, and the broad envelope corresponds to the phase noise. The pump and LO linewidths are found by fitting the top of the noise envelopes with Lorentzian curves [insets, Fig. \ref{figSelfheterodyne}(b) and (c)]. We extract a LO full-width-at-half-maximum (FWHM) linewidth of 223 kHz, which is close to twice the linewidth of the pump lasers (114 kHz). This corresponds to a LO coherence length of 428 m (200 m) in free space (waveguide) which is sufficient for nearly all continuous-variable (CV) quantum information protocols. 

The intensity noise of the LO is characterized directly by splitting the LO on a 50/50 beam splitter and detecting the outgoing beams with a balanced photodetector. We characterize the total noise of the LO by measuring the intensity noise on each photodiode separately. A balanced detection rejects the classical noise by $>$ 30 dB and retains the shot noise corresponding to the combined beam power. Due to the limited dynamic range of our balanced detector, the LO power is attenuated (375 $\mu$W) to avoid saturation of the single PD output, resulting in a RF noise close to the detector noise floor. We independently measure the noise floor of the detector and remove it numerically from LO noise spectra. As can be seen in Fig. \ref{figSelfheterodyne}(b), the classical noise of the LO extends beyond 50 MHz, which is the 3-dB gain bandwidth of our detector. This is attributed to the broadband thermal refractive noise present in silicon nitride \cite{LeThomas_Optica_2018,Huang_PRA_2019}. The balanced detection rejects these classical noise fluctuations, and we observe a noise spectrum matching the detector gain spectrum beyond 15 MHz. With the availability of ultrahigh-$Q$ microresonators \cite{Ji_Optica_2017}, the classical noise can be further reduced via integrated ring filters. In our experiments, we perform RF analysis around 40 MHz where the classical noise is sufficiently low and the detector gain is high.

\section{Theory of squeezed state generation}
\begin{figure*}
	\centering
	\includegraphics{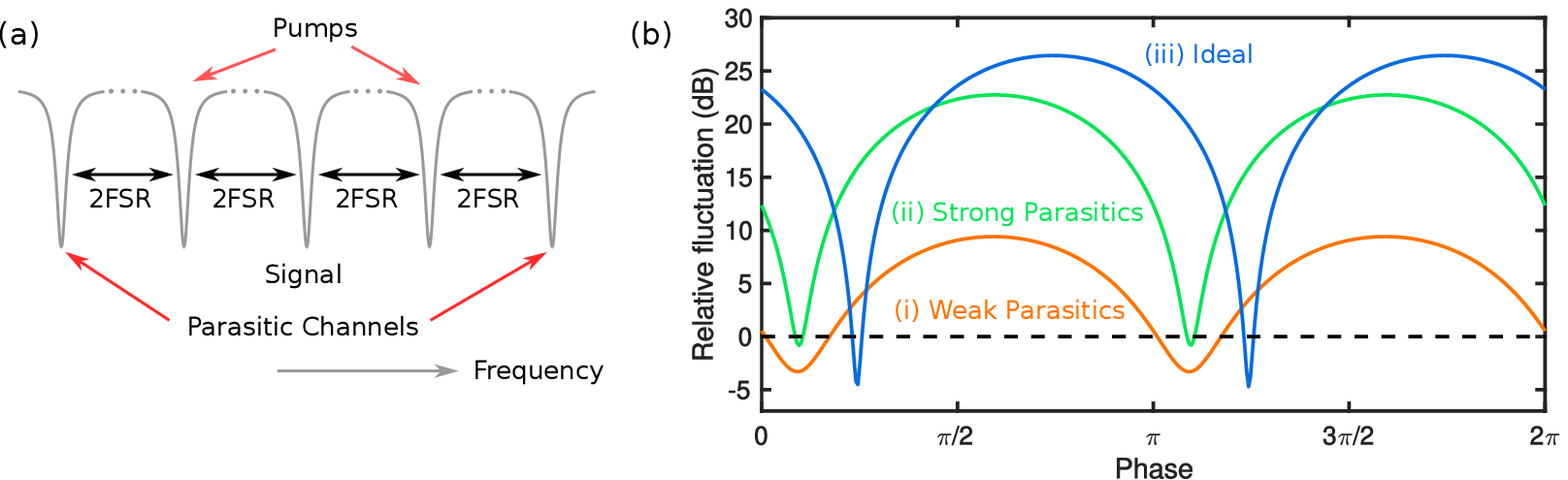}
	\caption{(a) Locations of the relevant resonances in frequency domain. (b) Theoretical calculation of squeezing generated by DPFWM, with parameters corresponding to the experiment (i), maximum parasitic coupling (ii), and no parasitic coupling (iii).}
	\label{figTheory}
\end{figure*}

A theoretical analysis of squeezed state generation with dual-pumped degenerate four-wave mxing (DPFWM) was shown in \cite{Vernon_PRAppl_2019}, where the analysis did not include potential parasitic processes. To understand the results of the current experiment, those processes must be included and the resulting model is presented in this section. There are 5 cavity modes that are most relevant to the experiment, the two pump modes, the signal mode and two dephasing channels due to parasitic nonlinear processes [Fig. \ref{figTheory}(a)]. In the small normal group-velocity-dispersion (GVD) resonator, the strengths of different nonlinear processes are determined by the interplay between SPM, XPM and pump detunings. Quantitatively, we define the pump detunings $\Delta_{-1}$ (red pump), $\Delta_1$ (blue pump) which include SPM and XPM effects as,
\begin{align}
&\Delta_{-1} = \delta_r - \gamma Lf_\delta(P_r+2P_b),\\
&\Delta_1 = \delta_b - \gamma Lf_\delta(2P_r+P_b),
\end{align}
where $\delta_r$ ($\delta_b$) is the red (blue) pump detuning with respect to the linear cavity, $\gamma$ is the nonlinear coefficient, $L$ is the cavity length, $f_\delta$ is the FSR, $P_r$ ($P_b$) is the red (blue) pump power inside the cavity. These definitions are the same as those in the main text. Similarly, the detuning of the signal mode can be found as,
\begin{align}\label{eqSignalDetuning}
\Delta_0 = \frac{\delta_r+\delta_b}{2} + 2\pi^2\beta_2 Lm^2f_\delta^3 - 2\gamma Lf_\delta(P_r+P_b),
\end{align}
where $\beta_2$ is the GVD coefficient, $m$ is the mode index difference between the pump and the signal modes. The first, second, and third terms correspond to the frequency tuning, dispersion effect, and XPM, respectively. The squeezing level is reduced if the central mode strongly couples to other modes through spurious nonlinear processes such as spontaneous four-wave mixing (SFWM) and four-wave mixing Bragg scattering (FWMBS). The detuning of these modes [Fig. \ref{figTheory}(a)] can be found as,
\begin{align}
&\Delta_{-2} = \frac{3\delta_r-\delta_b}{2}- 6\pi^2\beta_2 Lm^2f_\delta^3 -2\gamma Lf_\delta (P_r+P_b),\\
&\Delta_{2} = \frac{3\delta_b-\delta_r}{2}- 6\pi^2\beta_2 Lm^2f_\delta^3 -2\gamma Lf_\delta (P_r+P_b),
\end{align}
where the three terms have the same physical meaning as those in Eqn. (\ref{eqSignalDetuning}). In our low GVD and small mode separation system, the detuning term caused by dispersion is much smaller than the other terms and the linewidth. The parasitic nonlinear processes can be suppressed by creating significant detunings to $\Delta_{-2}$ and $\Delta_2$ while keeping a low detuning of $\Delta_0$. We choose parameters such that $\delta_r \approx 4\gamma Lf_\delta(P_r+P_b)$ and $\delta_b \approx 0$, i.e. a red detuned long-wavelength pump and small detuned short-wavelength pump with respect to the linear cavity resonances. A quick estimation shows that we can achieve a near-zero $\Delta_0$, a large positive $\Delta_{-2}$ and a large negative $\Delta_2$, indicating a suppression of the parasitic processes. The corresponding configuration is shown in Fig. 2(b).

To quantitatively estimate the effect of the parasitic processes, we first write down the equation of motion for each cavity mode annihilation operator \cite{Collett_PRA_1984,Vernon_PRA_2015,Zhao_Optica_2020}, which all have the same general form as,
\begin{align}
\frac{d\hat{a}(t)}{dt} = -\frac{i}{\hbar}[\hat{a}(t),\hat{H}_{sys}(t)]-\frac{\theta}{2}\hat{a}(t)-\frac{\alpha}{2}\hat{a}(t) - \sqrt{\theta}\hat{a}_{in}(t)-\sqrt{\alpha}\hat{b}_{in}(t),
\end{align}
where $\hat{H}_{sys}(t)$ is the cavity Hamiltonian, $\theta$ and $\alpha$ are the bus-ring coupling rate and photon scattering rate as shown in the main text, $\hat{a}_{in}(t)$ and $\hat{b}_{in}(t)$ are the annihilation operators corresponding to the cavity dissipation channels of bus waveguide and radiative modes, respectively. We assume the two pumps are classical undepleted fields, which yields a cavity Hamiltonian,
\begin{align}
&\hat{H}_{sys}(t) = \hat{H}_{sq}(t) + \hat{H}_{xpm}(t) + \hat{H}_{sf}(t) + \hat{H}_{bs}(t),\\
&\hat{H}_{sq}(t) = \hbar \gamma Lf_\delta \sqrt{P_rP_b}e^{-i(\omega_r t + \omega_b t + \phi_r + \phi_b)}\hat{a}_0^\dagger(t)\hat{a}_0^\dagger(t) + H.c.,\\
&\hat{H}_{xpm}(t) = 2\hbar \gamma Lf_\delta(P_r+P_b)[\hat{a}_0^\dagger(t)\hat{a}_0(t)+\hat{a}_{-2}^\dagger(t)\hat{a}_{-2}(t)+\hat{a}_{2}^\dagger(t)\hat{a}_{2}(t)],\\
&\hat{H}_{sf}(t) = \hbar\gamma Lf_\delta [P_re^{-i(2\omega_r t+2\phi_r)}\hat{a}_0^\dagger\hat{a}_{-2}^\dagger(t) + P_be^{-i(2\omega_b+2\phi_b)}\hat{a}_0^\dagger\hat{a}_{2}^\dagger(t)] + H.c.,\\
&\hat{H}_{bs}(t) = 2\hbar\gamma Lf_\delta \sqrt{P_rP_b}e^{-i(\omega_r t -\omega_b t+\phi_r-\phi_b)}[\hat{a}_0(t)\hat{a}_{-2}^\dagger(t) + \hat{a}_0^\dagger(t)\hat{a}_2(t)] + H.c.,
\end{align}
where $\hat{a}_0$, $\hat{a}_{-2}$ and $\hat{a}_2$ are the annihilation operators for the signal mode and two nonlinear dephasing channels respectively, $\hat{H}_{sq}(t)$, $\hat{H}_{xpm}(t)$, $\hat{H}_{sf}(t)$ and $\hat{H}_{bs}(t)$ are the Hamiltonians for DPFWM, XPM, SFWM and FWMBS respectively, $\phi_r$ and $\phi_b$ are the initial phases of the two pump fields which, we now show, can be eliminated through variable transformations. For the analysis of squeezing, we use the equation of motion on frequency domain operators \cite{Collett_PRA_1984} by performing the following transformations,
\begin{align}
&\hat{a}_0(\omega) = \frac{1}{\sqrt{2\pi}}\int_{-\infty}^{\infty}e^{i(\omega+\frac{\omega_r+\omega_b}{2})t+\frac{\phi_r+\phi_b}{2}}\hat{a}_0(t)dt,\label{eqTransform1}\\
&\hat{a}_{-2}(\omega) = \frac{1}{\sqrt{2\pi}}\int_{-\infty}^{\infty}e^{i(\omega+\frac{3\omega_r-\omega_b}{2})t+\frac{3\phi_r-\phi_b}{2}}\hat{a}_{-2}(t)dt,\label{eqTransform2}\\
&\hat{a}_2(\omega) = \frac{1}{\sqrt{2\pi}}\int_{-\infty}^{\infty}e^{i(\omega+\frac{3\omega_b-\omega_r}{2})t+\frac{3\phi_b-\phi_r}{2}}\hat{a}_2(t)dt,\label{eqTransform3}
\end{align}
where we also extracted carrier frequencies and chose convenient phase references. Because of the high $Q$ nature of the microresonators, each cavity mode only couple to a narrow frequency band of the input field operators $\hat{a}_{in}(t)$ and $\hat{b}_{in}(t)$. We can also perform a similar transformation as Eqn. [\ref{eqTransform1}-\ref{eqTransform3}] to these operators,
\begin{align}
&\hat{o}_0(\omega) = \frac{1}{\sqrt{2\pi}}\int_{-\infty}^{\infty}e^{i(\omega+\frac{\omega_r+\omega_b}{2})t+\frac{\phi_r+\phi_b}{2}}\hat{o}(t)dt,\\
&\hat{o}_{-2}(\omega) = \frac{1}{\sqrt{2\pi}}\int_{-\infty}^{\infty}e^{i(\omega+\frac{3\omega_r-\omega_b}{2})t+\frac{3\phi_r-\phi_b}{2}}\hat{o}(t)dt,\\
&\hat{o}_2(\omega) = \frac{1}{\sqrt{2\pi}}\int_{-\infty}^{\infty}e^{i(\omega+\frac{3\omega_b-\omega_r}{2})t+\frac{3\phi_b-\phi_r}{2}}\hat{o}(t)dt,
\end{align}
where $\hat{o} = \hat{a}_{in}, \hat{b}_{in}$. We now define vector operators,
\begin{align}
&\bm{\hat{r}}(\omega) = \left(\hat{a}_0(\omega),\hat{a}_0^\dagger(-\omega),\hat{a}_{-2}(\omega), \hat{a}_{-2}^\dagger(-\omega),\hat{a}_2(\omega), \hat{a}_2^\dagger(-\omega)\right)^\intercal,\\
&\bm{\hat{r}}_{in}(\omega) = \left(\hat{a}_{in,0}(\omega),\hat{a}_{in,0}^\dagger(-\omega),\hat{a}_{in,-2}(\omega), \hat{a}_{in,-2}^\dagger(-\omega),\hat{a}_{in,2}(\omega),\hat{a}_{in,2}^\dagger(-\omega)\right)^\intercal,\\
&\bm{\hat{u}}_{in}(\omega) = \left(\hat{b}_{in,0}(\omega),\hat{b}_{in,0}^\dagger(-\omega),\hat{b}_{in,-2}(\omega), \hat{b}_{in,-2}^\dagger(-\omega),\hat{b}_{in,2}(\omega), \hat{b}_{in,2}^\dagger(-\omega)\right)^\intercal.
\end{align}
All three vectors follow the commutation relation,
\begin{align}\label{eqCommutation}
[\bm{\hat{o}}(\omega),\bm{\hat{o}}^\intercal(\omega')]&= \bm{\hat{o}}(\omega)\bm{\hat{o}}^\intercal(\omega')-[\bm{\hat{o}}(\omega)\bm{\hat{o}}^\intercal(\omega')]^\intercal\notag\\
&=\begin{pmatrix}
0&1&0&0&0&0\\
-1&0&0&0&0&0\\
0&0&0&1&0&0\\
0&0&-1&0&0&0\\
0&0&0&0&0&1\\
0&0&0&0&-1&0
\end{pmatrix}\delta(\omega-\omega'),
\end{align}
where $\bm{\hat{o}}(\omega) = \bm{\hat{r}}(\omega),\bm{\hat{r}}_{in}(\omega),\bm{\hat{u}}_{in}(\omega)$. Now the equation of motion for the vector operator can be concisely written as,
\begin{align}
-i\omega\bm{\hat{r}} = i\bm{A}\bm{\hat{r}} -\frac{\theta+\alpha}{2}\bm{\hat{r}} - \sqrt{\theta}\bm{\hat{r}}_{in}-\sqrt{\alpha}\bm{\hat{u}}_{in},
\end{align}
where,
\begin{align}
&\bm{A} = \begin{pmatrix}
-\Delta_0 & \xi & \xi & \xi_r & \xi & \xi_b\\
-\xi & \Delta_0 & -\xi_r & -\xi & -\xi_b & -\xi\\
\xi & \xi_r & -\Delta_{-2} & 0 & 0 & 0\\
-\xi_r & -\xi & 0 & \Delta_{-2} & 0 & 0\\
\xi & \xi_b & 0 & 0 & -\Delta_{2} & 0\\
-\xi_b & -\xi & 0 & 0 & 0 & \Delta_{2}
\end{pmatrix}\\
&\xi = 2\gamma Lf_\delta\sqrt{P_rP_b},\\
&\xi_r = \gamma Lf_\delta P_r,\\
&\xi_b = \gamma Lf_\delta P_b.
\end{align}
This linear system can be readily solved with a matrix inversion. With the input-output relation \cite{Collett_PRA_1984}, we can express the output field as,
\begin{align}
\bm{\hat{r}}_{out}(\omega) = (\bm{I}+\theta\bm{M})\bm{\hat{r}}_{in} + \sqrt{\theta\alpha}\bm{M}\bm{\hat{u}}_{in},
\end{align}
where $\bm{I}$ is the identity matrix and,
\begin{align}
\bm{M} = [i(\omega\bm{I}+\bm{A})-\frac{\theta+\alpha}{2}\bm{I}]^{-1}.
\end{align}
$\bm{\hat{r}}_{out}(\omega)$ together with the commutation relation Eqn. (\ref{eqCommutation}) contain all the statistical information of the output fields. By defining,
\begin{align}
\bm{\Psi} = \begin{pmatrix}
1 & e^{-i2\psi} & 0 & 0 & 0 & 0\\
e^{i2\psi} & 1 & 0 & 0 & 0 & 0\\
0 & 0 & 0 & 0 & 0 & 0\\
0 & 0 & 0 & 0 & 0 & 0
\end{pmatrix},
\end{align}
we can write the quadrature fluctuation of the signal field as,
\begin{align}
S(\psi; \omega)\delta(\omega-\omega^\prime) = \braket{\bm{\hat{r}}_{out}^\dagger(\omega)\bm{\Psi}\bm{\hat{r}}_{out}(\omega^\prime)},
\end{align}
We further let $\bm{S} = (\bm{I}+\theta\bm{M}^\dagger)\bm{\Psi}(\bm{I}+\theta\bm{M})+\theta\alpha\bm{M}^\dagger\bm{\Psi}\bm{M} = (s_{ij})$, then,
\begin{align}
S(\psi; \omega) = s_{22}+s_{44}+s_{66},
\end{align} 
i.e. the quadrature fluctuation can be calculated as the sum of the (2,2), (4,4) and (6,6) elements of the numerical matrix $\bm{S}$. 

Since only two cavity modes are responsible for the degradation of the squeezed state, we can use a dual ring structure \cite{Little_JLT_1997} to drastically change the local dispersion of those two modes without influencing the pump and squeezed modes. For example, since our current experiment utilize pumps that are 2-FSR away from the squeezed mode, a secondary ring with a radius $\frac{n}{8}$ (n is any small odd number) of the main ring can create large mode interactions on the two undesired cavity modes, leaving only the squeezing process phase-matched (together with the SPM and XPM processes). Such a structure allows us to realize $\Delta_{-2}, \Delta_2 \gg \xi, \xi_r, \xi_b$, which reduces $S(\psi; \omega)$ to the ideal squeezing case with a minimum fluctuation given by,
\begin{align}\label{eqIdealSqueezing}
S(\psi; \omega)_{\text{min}} = S(\frac{\pi}{4}; 0) = \frac{(\xi-\frac{\theta-\alpha}{2})^2+\theta\alpha}{(\xi+\frac{\theta+\alpha}{2})^2}, 
\end{align}
and an oscillation threshold given by $\xi = (\theta+\alpha)/2$. Equation (\ref{eqIdealSqueezing}) is exactly the same as one would get from a $\chi^{(2)}$ based degenerate parametric down conversion process. 

We estimate the parameters in our experiment based on the calculation shown in Fig. 2(b), which yields $\delta_r = 2\pi\times 650$ MHz, $\delta_b = 0$ MHz, $P_r =$ 9.5 W, $P_b$ = 3.2 W, $\theta = 2\pi\times 300$ MHz, and $\alpha = 2\pi\times 150$ MHz. These values lead to $\xi = 2\pi\times 220$ MHz, $\xi_r = 2\pi\times 200$ MHz, $\xi_b = 2\pi\times 60$ MHz, $\Delta_0 = 2\pi\times 165$ MHz, $\Delta_{-2} = 2\pi\times 355$ MHz, and $\Delta_{2} =2\pi\times  (-945)$ MHz. The corresponding quadrature fluctuation $S(\psi; 0)$ is shown in Fig \ref{figTheory}(b)(i) with a maximum noise reduction of 3.5 dB. As a comparison, we calculate the squeezing level with maximum nonlinear parasitic coupling by setting $\Delta_0 = \Delta_{-2} = \Delta_2 = 0$ MHz, which results in a squeezing level of only 0.8 dB [Fig. \ref{figTheory}(b)(ii)]. We also note that in this case higher squeezing levels can be achieved with lower pump powers. The ideal squeezing is calculated by setting $\Delta_0 = 0$ MHz, $\Delta_{-2} = 2\pi\times 3550$ MHz, $\Delta_{2} =2\pi\times  (-9450)$ MHz and we get the ideal quadrature fluctuation for the current cavity coupling condition with a squeezing level of 4.7 dB [Fig. \ref{figTheory}(b)(iii)].

\section{Squeezing data analysis}
Due to technical issues such as mechanical vibrations and radio frequency spectrum analyzer measurement time, both the scan duration and the scan resolution are limited, which limits the number of data points in each scan. We estimate the measured squeezing level as follows. In each valley, we take the data point at the lowest noise level and the two data points next to it, resulting in 18 data points from 6 valleys. We then remove an anomalous point caused by optical component vibration in the measurement system. The squeezing level is calculated with the remaining 17 data points, which could cause an underestimation of the actual squeezing level as the 3 data points in each valley cover a 12$^\circ$ phase interval around the squeezed quadrature, which effectively introduces a portion of the anti-squeezed quadrature noise into the measurement.

Let $V_d$ be the detector dark noise variance and $V_{ph}$ be the squeezed photocurrent noise variance, the directly measured squeezed current variance is subsequently $V=V_d+V_{ph}$. Additionally, let $V_x$ be the squeezed quadrature fluctuation without loss, $V_s$ be the LO shot noise variance, and $\eta$ be the detection efficiency, $V_{ph}$ can be expressed as $V_{ph} = \eta V_x+(1-\eta)V_s$. As we are only concerned with the ratios of the variances, we express all variances in terms of microwave powers acquired by the RFSA. From traces (iii) and (ii) in Fig. 4(b) in the main text, $V_d=13.5$ pW and $V_d+V_s=37.1$ pW. As both traces have sufficient data points and small fluctuations, we treat these two values as exact. From trace (i) in Fig. 4(b) in the main text, $V_{ph} = 30.8 (2.7)$ pW, where the value in the parenthesis represents the standard error of the dataset defined as $\text{(standard deviation)}/\sqrt{\text{sample size}}$. The directly observed noise squeezing can be calculated as $(V_{ph}+V_d)/(V_s+V_d) = 0.83 (0.018) = -0.81 (0.09)$ dB where the standard error on the dB scale is defined as 10log$_{10}$[(standard error + mean)/mean]. After correcting for detector dark noise, we get $V_{ph}/V_s = 0.73 (0.018) = -1.34 (0.16)$ dB. After correcting for the losses $\eta = 52\%$, we infer the generated squeezing level $V_x/V_s = 0.49 (0.05) = -3.09 (0.49)$ dB.

\section{A path to CV quantum computing}
CV quantum information processing can find promising applications in both quantum simulation such as molecular vibronic spectroscopy \cite{Huh_NatPhot_2015} and universal quantum computing \cite{Menicucci_PRL_2014}. There are 3 key hardware ingredients to implement most of the quantum simulation and computation protocols, namely sources of squeezed states, large-scale low-loss interferometric networks, and high-efficiency detection schemes.

This work is aiming to provide a pathway to the first ingredient. The ideal squeezed states should be in single spatiotemporal modes and have high squeezing levels \cite{Vernon_PRAppl_2019}. In our on-chip scheme, a single spatial mode is easily achieved by exciting both pump fields in the fundamental waveguide modes. Due to the large modal dispersion, both the local oscillator and the squeezed state are produced only in the fundamental modes. A single temporal mode can be achieved by employing pulsed pump fields that match the cavity linewidth, as shown theoretically in \cite{Vernon_PRAppl_2019}. Fault-tolerant quantum computing requires more than 10 dB of squeezing \cite{Fukui_PRX_2018}. In our generation scheme, this demands the intrinsic $Q$ to be more than 10 times larger than the loaded $Q$. As the current squeezing ring has a loaded $Q$ of $4.3\times10^5$, improving the intrinsic $Q$ from $1.3\times10^6$ to $4.3\times10^6$ can produce $\approx$ 10 dB squeezing with current pump powers. While silicon nitride microresonators with intrinsic $Q$ of $6.7\times10^7$ have previously been reported \cite{Ji_Optica_2017}, it is highly dependent on the waveguide crosssection dimension. The achievable $Q$ within the current technology limit for the waveguide dimensions used in this work is still under investigation.

On-chip interferometer networks are key components for both DV and CV quantum information processing. The main loss mechanism in such networks is propagation loss, which is as low as 0.08 dB/cm on the SiN platform. Large interferometer arrays have been now been routinely implemented in quantum photonic experiments \cite{Carolan_Science_2015}.

Two types of detection schemes are commonly used for CV related applications, homodyne/heterodyne detection and photon-number-resolving (PNR) detection, which have both been demonstrated on-chip \cite{Calkins_OE_2013,Raffaelli_QST_2018,Porto_JOSAB_2018}. Since the goal of universal quantum computing is to directly measure at least 10 dB of squeezing, the detection loss should be well within 10\%. Waveguide based photodetectors have achieved 90\% quantum efficiency \cite{Yu_OE_2018} at telecom wavelengths with higher efficiencies theoretically possible \cite{Piprek_SPIE_2003}. Standalone detectors have shown 97\% quantum efficiency at telecom wavelengths \cite{Lin_PhotRes_2017} and 99.5\% at near-infrared wavelengths \cite{Vahlbruch_PRL_2016}. With high efficiency homodyne detections, all Clifford gates can be implemented for quantum computing \cite{Menicucci_PRL_2014}. PNR detection is important for Gaussian boson sampling and can provide the non-Clifford gates in quantum computing. While homodyne detection is highly tolerant to noise photons far away from the LO frequency, PNR detector requires a more stringent rejection of noise-photons, particularly those from the strong pump fields. It has been shown that such noise photon rejection can be adequately achieved by on-chip Bragg filters \cite{Oser_arXiv_2020}. Additionally, it is worth mentioning that there exist CV quantum computing schemes that does not require the use of PNR detections \cite{Baragiola_PRL_2019}. With the help of Gottesman-Kitaev-Preskill encoding, homodyne/heterodyne measurements are sufficient for universal quantum computing.

\end{document}